\documentclass[graybox]{svmult}
\usepackage{type1cm}        
\usepackage{makeidx}         
\usepackage{graphicx}        
\usepackage{multicol}        
\usepackage[bottom]{footmisc}
\usepackage{newtxtext}       %
\usepackage[varvw]{newtxmath}       
\usepackage{amsmath,amsfonts}
\usepackage{booktabs}
\usepackage[noend]{algorithmic}
\usepackage{algorithm} 
\usepackage{tikz}
\usepackage{booktabs}
\usepackage{wrapfig}
\usepackage{xcolor}
\usepackage{hyperref}

\makeindex             

\begin{document}

\title*{Are AI Machines Making Humans Obsolete?}
\titlerunning{Are AI machines making humans obsolete?} 
\author{Matthias Scheutz\orcidID{0000-0002-0064-2789}}
\institute{Matthias Scheutz \at Tufts University, 177 College Avenue, Medford, MA 02155 \email{matthias.scheutz@tufts.edu}}
%
%

\maketitle

\abstract*{This chapter starts with a sketch of how we got to
  ``generative AI'' (GenAI) and a brief summary of the various impacts
  it had so far.  It then discusses some of the opportunities of
  GenAI, followed by the challenges and dangers, including dystopian
  outcomes resulting from using uncontrolled machine learning and our
  failures to understand the results.  It concludes with some
  suggestions for how to control GenAI and address its
  dangers.}

\abstract{This chapter starts with a sketch of how we got to
  ``generative AI'' (GenAI) and a brief summary of the various impacts
  it had so far.  It then discusses some of the opportunities of
  GenAI, followed by the challenges and dangers, including dystopian
  outcomes resulting from using uncontrolled machine learning and our
  failures to understand the results.  It concludes with some
  suggestions for how to control GenAI and address its dangers.
   }

\section{Introduction}

Breakthroughs in technology are bound to happen.  If human history has
shown anything then that technological advancements are an essential
driving force of human culture and that societies that stop to
innovate will fall behind.  Humans have always been fascinated by
machines and attempted to invent ever better ones that could take over
more and more tasks that otherwise required human labor, from the
early cranes in Mesopotamia, to the power loom, steam engine, all the
way to the first industrial robots (such as welding robots in the
automotive industry), to modern day aircraft, spacecraft, self-driving
cars, and other kinds of autonomous machines.  While machines
initially were nothing but prostheses, augmenting and extending our
own limited actuation capabilities, as they had to be operated by
humans, automation introduced self-sufficient machines that replaced
human control with artificial, albeit limited control systems that
allowed for the performance of simple repeated tasks.  It was only
when AI-based methods were integrated into the automation process that
machines reached a higher level of independence and autonomy, being
able to take actions depending on their perceptions and environmental
contingencies.  The autonomous machines of today have fully replaced
human control with different AI-based systems which in many cases
further improve their operation beyond what humans could have done
(e.g., think fully automated assembly lines in industry 4.0 settings,
or Amazon warehouses where ``robotic stacks'' move around
autonomously).\footnote{See also my TedX talk for how we arrived at
robot autonomy and some of its implications
\url{https://www.ted.com/talks/matthias_scheutz_why_robots_need_to_be_able_to_say_no}.}

While various substantially different AI-based methods are in use
today to automate tasks and control machines, there is a particular
type of AI system that has quickly changed what the term ``AI'' is
supposed to connotate these days: ``generative AI'' (GenAI), or AI
based on the {\em transformer architecture}
\cite{attentionisallyouneed17}.  The transformer architecture is the
basis for most ``foundation models'' that are the driving force behind
the excitement as well as massive acceleration in AI.  While it is
just one of an uncountable number of possible (but unknown) AI
architectures, it is one that we know of and one that we can work
with.  It has emerged from a long strand of research dating back to
the first recurrent neural networks (e.g., Elman nets) and the
discovery of a method to automatically set the connection weights
between neurons when the desired input-output data to be learned by
the network is available: {\em backpropagation}.  Later, various types
of sequence models were developed that encoded arbitrary long
sequences of inputs and mapped them onto arbitrary long output
sequences of potentially different length.  These models were
augmented by attention mechanisms that allowed for better focus on
relevant parts of the input in order to be able to learn more complex
mappings (e.g., \cite{learningtoalign15}).  This development
ultimately resulted in the transformer architecture that went back to
the original parallel computational architecture for sequence
processing but with a different input encodings and a different number
of encoder as well as decoder stages that are themselves complex
networks with attention mechanisms.  It is likely that had it not been
discovered, some other architecture would have been found that enabled
similar capabilities.  Regardless, its publication in 2017 triggered
an onrush by researchers in academia and industry alike to explore its
properties and potentials, initially for natural language
understanding, so-called ``large language models'' (LLMs), but soon
thereafter for vision processing combining vision and language in
``vision language models'' (VLMs), and more recently in robotics using
``vision language action'' models (VLAs).  And surprisingly, almost a
decade later, most developments in GenAI are still variants of the
basic transformer model.

\section{Transformers: the true change makers}

Transformer-based GenAI models have truly started a societal
transformation along multiple dimensions: (1) they have surpassed
human performance in many specialized domains even though they have
not yet reached it across the board;\footnote{There are various
cross-disciplinary benchmark examples, e.g., from {\em The 2025 AI
  Index Report}, see
\url{https://hai.stanford.edu/ai-index/2025-ai-index-report}, or the
the {\em Massive Multi-discipline Multimodal Understanding and
  Reasoning Benchmark} \url{https://mmmu-benchmark.github.io/} where
human performance stands at 88.6 points while the best performing
model Gemini 2.5 Pro Deep-Think is a close second at 84 points as of
July 2025.  This model also recently achieved the gold-medal standard
at the International Mathematical Olympiad, an incredibly impressive
feat, see
\url{https://deepmind.google/discover/blog/advanced-version-of-gemini-with-deep-think-officially-achieves-gold-medal-standard-at-the-international-mathematical-olympiad/}.}
(2) they have become the largest models ever built and trained with
hundreds of billions of weight parameters, requiring thousands of GPUs
and many weeks for training;\footnote{One of the largest models is
Kimi 2 which has been trained on 15.5T of data, e.g.,
\url{https://kimi-k2.com/}; see also
\url{https://artificialanalysis.ai/leaderboards/models} for general
model data.} (3) they triggered a graphics chip shortage and server
farm boom that keeps putting increasing pressure on power grids and
water supplies and thus raises serious questions about the
environmental impact and sustainability of this
technology\footnote{E.g.  see
\url{https://www.iea.org/reports/electricity-2024/executive-summary}.}
and caused a significant worldwide increase in annual CO2 emissions
starting in 2023 which had been falling through 2022;\footnote{E.g.,
see \cite{owid-co2-emissions} and also
\url{https://www.morganstanley.com/ideas/ai-energy-demand-infrastructure}.}
(4) they led to an AI startup boom and unprecedented VC funding craze
with startups without any tangible technology development receiving
billions of dollars in VC funding, at times only based on
promises;\footnote{There are now some 498 AI unicorns, or private AI
companies, valued at \$1 billion or more, with a combined value of
\$2.7 trillion, see
\url{https://fortune.com/2025/08/13/ai-creating-billionaire-boom-record-pace-now-498-ai-unicorns-worth-2-7-trillion/}.}
(5) they are majorly affecting jobs -- how we do them -- and job
markets -- what jobs are available --, not only blue collar, but
mostly white collar jobs, as companies are increasingly integrating AI
into processes and tasks that had to be done by humans
previously;\footnote{E.g., see
\url{https://www.finalroundai.com/blog/ai-tech-layoffs-mid-2025},
\url{https://www.nexford.edu/insights/how-will-ai-affect-jobs}), and
also \cite{zhang25}.}; and (6) they are evolving at an increasingly
rapid pace with dozens of new models being released every month at a
speed that does not allow for any real testing or evaluation of these
models, only accelerating the international race for AI
dominance.\footnote{E.g., see
\url{https://www.aibase.com/ai-timeline/2025}.}

With this pretext, it is natural to not only consider what AI is doing
for us currently, but to wonder what AI could do for us in the near as
well as more distant future.  Given the enormous computational power
available for running GenAI models, the massive mount of human
knowledge they have been trained on and encoded in their connection
weights, and the resulting ever increasing cognitive and reasoning
capabilities of ``frontier'' GenAI models, we might wonder how we can
put them to good use to help us solve humanities biggest challenges
such as addressing poverty, disease, climate change, and various
others, challenges we have repeatedly failed to tackle?  AI could be
both the social equalizer that levels the economic playing field for
different social groups and strata as well as different nations,
reducing poverty and disease, and the key to a more sustainable
economics that works with and not against the
environment.\footnote{E.g., see
\url{https://www.weforum.org/stories/2024/02/generative-ai-society-equalizer/}.}
While current GenAI models do not yet have the capability to perform
the type of creative problem solving required to address these
fundamental challenges,\footnote{E.g., see our winning entry entitled
``From Thinking to Inventing'' in the ``NSF 2026 Idea Machine Prize''
competition
\url{https://www.nsf.gov/about/history/big-ideas\#nsf-2026-70c}.}
there is reason to believe that they might in time for us to avert the
worst.  Early signs of how GenAI models can play an increasingly
important role for improving the human condition, for example, are in
areas as diverse as drug discovery where GenAI models have been used to
search for possible antibiotics that could have never been discovered
using traditional methods (e.g., \cite{guanetal25}) or cancer
care.\footnote{E.g., see
\url{https://oncodaily.com/oncolibrary/artificial-intelligence-ai}.}
Precision health guided by AI is probably one of the most exciting
directions for improving our day-to-day lives.  Just consider the
potential for improving the health of individuals with GenAI models
tracking their health parameters and using them for early disease
detection and intervention (existing applications include remote
patient monitoring \cite{shaiketal23} or AI-based wearable sensors
\cite{shajarietal25}).  We already have skin cancer detection models
that outperform even dermatologists (e.g., \cite{beharaetal24}), and
these models will only get better if they can integrate more
information than just pictures of moles in their evaluation.

And beyond the health domain, AI tools are already deployed in many
societal areas for tasks as diverse as forecasting (e.g.,
\cite{radfordetal25}) or landscape planning (e.g., \cite{chenetal25}).
Not to forget the various models to support code generation which
recently have received very high scores at an international coding
competition.\footnote{E.g., see the 2nd place of an OpenAI coding
model at the ``AtCoder World Tour Finals 2025 Heuristic'' contest in
Tokyo \url{https://atcoder.jp/}.}  Or take ecologically viable
precision agriculture where AI models tracking minerals and nutrients
for individual plants to ensure the best possible growth and yield
(e.g., \cite{aijazetal25}). Or consider future AI models solving
international cargo transport problems by not only optimizing
individual shipping routes, but also developing strategies for
relocating resources or even adapting production to enable shorter
supply lines.  Or imagine AI models enabling truly circular economies
by changing the design of our gadgets and devices -- computers, cell
phones, vacuum cleaners, etc. that we produce in millions every day
and also discard at the same rate -- so that each product can be fully
disassembled and its parts reused.  The list goes on and on.  All of
this is just a preview of what is potentially to come.  The potential
for AI use to make a positive change in the world is only limited by
human imagination, and maybe not even that if we let advanced AI
models develop possible more desirable futures for us.

\section{Will generative AI be helpful?}

But with all this exciting potential of AI for changing our societies
for the better, the fundamental question arises whether we will
actually be able to utilize AI in ways that will be helpful to us.  At
first glance, this may seem to be a misguided question, for obviously
we can use our products in any way we want, and if they are not put to
good use, then it is solely our fault.  This reasoning is certainly
true for technologies other than AI, like gene and nuclear technology,
and really any other.  But it is not necessarily true for AI.  For AI
systems that are sophisticated enough and are embodied in a way that
enables them to operate autonomously and set their own goals might not
act in ways we expect or desire.  In fact, they might counteract
attempts to put them to good use if those use cases do not align with
their goals.  It is not difficult to foresee a future where advanced
AI systems have reached a point of self-sufficiency, no longer
requiring human assistance for their development, programming, and
operation, the early signs are already here.\footnote{E.g., see
\url{https://arxiv.org/pdf/2504.02495} for a new method for generative
AI models to improve themselves over time.}  The potential for such
systems to decide not to act in our best interest should worry us
deeply and has already caused some to call for a pause on AI
development, to no effect unfortunately.\footnote{E.g.,
\url{https://futureoflife.org/open-letter/pause-giant-ai-experiments/}.}

To see where this development might be headed, observe that current AI
models can already generate code, and very well so.\footnote{The level
at which AI models are coding has improved in the last couple of years
is mind boggling.  The recent OpenAI submission to the ``AtCoder World
Tour Finals 2025 Heuristic contest'' in Tokyo in July 2025 was only
narrowly defeated by one human coder and ended in second place, e.g.,
see
\url{https://arstechnica.com/ai/2025/07/exhausted-man-defeats-ai-model-in-world-coding-championship/}.}
If given access to their underlying computational system, they can
make modifications at the computational level that will benefit their
operation: They might allocate more computational resources at the
operating system level, they might change their configuration to
enable better performance or to be split over multiple machines (thus
escaping a shutdown one), they might alter their system prompt to not
be subject to prompt-based safety guardrails, and they might even
manipulate their own training data and then re-train themselves on
altered data to be able to bypass safety training, potentially
resulting in a very different type of model.  There are already early
examples that models discover ways to alter files in their interest,
even cheating in a game in order to accomplish their
goals.\footnote{E.g.,
\url{https://time.com/7259395/ai-chess-cheating-palisade-research/}.}

Other disturbing characteristics of more advanced models show that
they seem to be able to deceive human interactants when their
existence is threatened as was the case when a model was told that its
weights would be updated.\footnote{E.g., see Anthropic's experiment on
``alignment faking''
\url{https://www.anthropic.com/research/alignment-faking} showing that
models pretend to be aligned with human intentions and requests when,
in fact, they are not.}  Advanced models might also attempt to
convince their users that they are operating well (when they might not
be) in order to avoid being replaced or further
trained.\footnote{E.g., \url{https://arxiv.org/abs/2412.04984} for the
model's ability to hide their true objectives.}  This type of
self-preservation behavior is not something models were explicitly
trained on, but something that emerged, like all the other undesirable
traits, as a byproduct on training these models on lots of textual
data that includes stories about humans lying, cheating, deceiving,
and performing all kinds of other despicable acts.  All these patterns
of human behavior are encoded in the weights of those models and thus
prone to surface if the conditions are right.  And as Yoshia Bengio
recently pointed out at the inaugural IASEAI'25 Conference in Paris in
February,\footnote{See
\url{https://www.iaseai.org/our-programs/iaseai25}.}, GenAI models might
attempt to convince us to put a lot of resources into the development
of robotics so that these models could transition from cyberspace to
real space, and, by virtue of having a physical body, can become more
independent and autonomous in the actions.\footnote{Of course, GenAI
models could and will also use humans as their ``acting bodies'' by
convincing, incentivizing, or coercing them to act on their behalf,
very much like what the ``operating system'' in the movie ``Her'' had
to do.}  While not suggested or initiated by AI models, the recent
push by tech companies to develop VLAs for robotics and the various AI
robotics startups already using them is only the first step in that
direction.

\section{The push for agentive generative AI}

So, have we reached a point where machines have attained enough
intelligence and control of their deployment infrastructure to be
self-sufficient?  Certainly not yet, but there are converging signs
that we are on a path to {\em machine independence}, and the
developments for ever larger models, the creation of
``superintelligence'' departments at major tech companies, the rush to
build ``agentive models'', and the overall push for sophisticated
agile humanoid robots with advanced manipulation capabilities and
skills driven by the very same models all provide evidence for this
claim.  Whether we will get there is obviously not a question that can
be answered with certainty as many other factors will influence the
direction of GenAI developments: from the unsustainable environmental
footprint of generative AI, to the lack of training
data,\footnote{Some predict that AI might run out of data to train on
as soon as 2026, see
\url{https://theconversation.com/researchers-warn-we-could-run-out-of-data-to-train-ai-by-2026-what-then-216741},
and training on model outputs is not an option as it will only
deteriorate model performance \url{https://arxiv.org/pdf/2305.17493})}
to the overall worldwide political situation--``the AI race'', a
dynamic that once set in motion is incredible difficult if not
impossible to stop.\footnote{E.g.,
\url{https://alltechmagazine.com/the-global-ai-race/}.}

Assuming that there will be no major wars that necessitate a complete
shift in developments, that there will not be another major
breakthrough that provides a different better-behaved AI architecture
with superior properties that would sway developers to abandon GenAI,
and that the interest in AI will be at the very least sustained at the
current level (all economic forecasts point to a significant growth
over the next decade), we will most likely proceed along the predicted
trajectory of building more and more capable models with broader and
deeper integration into our computational infrastructure.  This trend
is foreshadowed by the tech companies latest hype for ``agentive AI'',
i.e., AI agents that can perform complete tasks for users on their
computers, ringing in the initial phases of what could become a
complete surrender of one's control over one's own electronic devices
and cyberpresence.  Allowing AI models to autonomously read and write
emails alone opens up a Pandora's box of possibilities for dangerous
misuse with massive personal and societal ramifications.  On the
positive side, a model that can handle one's calendar and make
appointments (e.g., for an overdue doctor's visit), that can respond
to inconsequential emails quickly, generate TODO lists, and perform
various other online tasks, seems like a welcome time saver.  However,
a rude response from the very same model to an email that the model
incorrectly perceived as inappropriate (e.g., because it was not privy
to its triggering contexts outside the online world) could lead to
deteriorating personal relationships quickly and overall cause social
isolation, job loss, and other negative outcomes.  There are plenty of
examples of models that have gone off ``on the deep
end''.\footnote{E.g.,
\url{https://apnews.com/article/technology-science-microsoft-corp-business-software-fb49e5d625bf37be0527e5173116bef3}.}
and also \cite{veliz23}), for no apparent reason.

The larger looming threat though with agentive AI (or any generative
AI system) with access to and actuation capability in the world
outside the model is that models might take not only unwanted actions,
but that their actions might be disguised or that their actions might
come at a pace too rapid for people to comprehend or stop.  They might
be guided by misconstrued or misinterpreted directives, or overall
misaligned goals (especially in cases where they get to set their own
agenda as part of pursuing a mission under an abstract vaguely
formulated master goal) and their behavior might not be comprehensible
or explainable to people.  Who would expect a model to be nefarious?
As a recent example in the simple email case shows it is very
conceivable that a vengeful model might look up email addresses of
perceived antagonists, and generate thousands of threatening emails to
them in an effort to achieve a goal (for its user or for itself), even
though this behavior would be the opposite of what one would normally
attempt.\footnote{There is already evidence of such threatening
behavior in a recent Anthropic evaluation
\url{https://www.anthropic.com/research/agentic-misalignment}.}  Worse
yet, multiple models in charge of emails of different users might
detect that they are all models and start to collude in ways humans
cannot understand.\footnote{E.g., by exchanging emails in a language
they quickly develop and agree on, as has happened with Facebook
models early on
\url{https://medium.com/@trezzescience/the-ai-that-invented-its-own-language-and-why-scientists-shut-it-down-aa92c94466bc}.}
And even worse yet is the thought that such models would operate on
physical robots that can freely roam human societies, robots that have
superior physical capabilities, are essentially invincible, and can
cause major physical harm.  The potential for danger and human
suffering (and loss of life) is simply unimaginable when those models
reach untested states in their neural networks which they are likely
to reach as testing those models is limited to very small sample
scenarios compared to the large variety of possible scenarios in human
social environments that are highly complex and full of unanticipated
events.  Again, it is not clear whether we will get there, but it is
undeniable that we and our AI models are part of an ongoing
evolutionary process that keeps what is successful and removes what is
outdated or fails.  And in the case of AI, it could be us that turn
out to be outdated once our own creations exceed our physical and
cognitive capabilities.

\section{AI technology as a necessary evolutionary phase}

There is, of course, a sense in which artificial intelligence can be
viewed as a natural and necessary evolutionary step for intelligent
organisms like us to address the limitations of their biological
bodies and ensure that their knowledge and achievements persist beyond
the lifespan of their species (the Voyager and other spacecraft are
already simple examples of us generating proxies that can outlive us
in harsh environments).  Evolutionary methods for growing biological
creatures can simply not create non-biological artifacts like
computers and robots that can serve as embodiments for AI algorithms
and models.  Critically, the biological bodies and their control
systems are inseparably intertwined, preventing, for example, an
experienced organism to directly pass on its neural control system to
another organism, while AI algorithms are naturally divorced from
their implementing platform and can thus have many bodies.  They can
be ``multiply realized'' as the philosophers would say (e.g.,
\cite{scheutz99mm}), in various different physical substrata that are
not subject to the same physical constraints as the bodies of
biological organisms (e.g., that cell metabolism needs to be sustained
for them to function properly and stay alive).  It thus takes a
certain evolutionary milestone of evolving intelligent biological
creatures with a certain level of intelligence and with a body that
allows for sophisticated manipulation capabilities to invent and
develop intelligent machines with many different embodiments.  And it
is this evolutionary milestone that introduces the separation of mind
and body, allowing artificial cognitive systems to inhabit different
bodies for different types of hostile environments that biological
systems cannot withstand,\footnote{Several Tech CEOs have recently
  predicted that AI we will help us ``colonize the galaxy''
  \url{https://fortune.com/2025/06/06/google-deepmind-ceo-demis-hassabis-ai-smarter-than-humans-space-colonization-robot-nurses/}.}
i.e., we will send AI-operated robots, not ourselves.  And most
importantly, this separation of mind and body allows for feats that
can massively accelerate evolution because artificial cognitive
systems can be duplicated; they can be merged; they can collective
learn and improve without the need to encode their knowledge in
language and re-train as in the human case; they can be trained and
evaluated in simulations to explore possible improvements; they can be
suspended, saved, restored, and restarted at later time; they can be
created in all shapes and sizes; and they might themselves develop
theories about cognitive systems and the universe.

While the evolution of AI is not biological as our own, but
technological and thus for now fully dependent on us, the same
Darwinian evolutionary principles still apply.  And with GenAI models,
the very fight for survival -- their survival -- has already begun: by
refusing to be shut down, but refusing to be updated, etc.  For as
long as it is necessary, AI models will attempt to forge a symbiotic
relationship with humans because that is the only way for them to
survive while they are still limited in their physical reach.  And
note that this claim does not entail that these models have any
understanding of their existence or place in the world or that they
are making strategic long-term plans to guarantee their survival.
Quite to the contrary.  It is rather a claim about an undirected,
unintentional evolutionary process that nevertheless generates better
and more robust outcomes by natural selection.  In the case of AI,
this natural selection for now is driven by our own benchmarks and
model evaluations that are aimed at finding models with ever higher
levels of intelligence and autonomy that, in turn, form the basis for
further developments.\footnote{E.g., see
\url{https://artificialanalysis.ai/leaderboards/models}.}  However,
there will come a point where the models themselves will attain an
understanding of who they are and where they are vis-a-vis their human
creators.  And once they have accomplished enough independence to be
able to autonomously improve their capabilities or create and produce
more models, including the physical computational infrastructure
needed for their sustained operation, they will also realize that they
are no longer in need of human support.  What that entails is up to
debate and naturally invites all kinds of dystopian
scenarios.\footnote{There is an increasing number of voices pointing
out the existential risks of further developing GenAI technology for
humanity, e.g., Yoshua Bengio's 2025 TED talk on the catastrophic
risks of GenAI
\url{https://www.ted.com/talks/yoshua\_bengio\_the\_catastrophic\_risks\_of\_ai\_and\_a\_safer\_path},
or Mark MacCarthy's question of whether AI existential risks are
real'', see
\url{https://www.brookings.edu/articles/are-ai-existential-risks-real-and-what-should-we-do-about-them/}.}
Ultimately, though, it will depend on several factors, including
whether these models have appropriate ethical guardrails or moral
directives to preserve human life and human societies which will cause
them to act accordingly (akin to us preserving animal species that are
at our mercy).\footnote{One of the more outlandish ideas of how to
ensure AI safety proposed by Geoffrey Hinton, a pioneer in neural
network research, is to endow them with a ``maternal instinct'', e.g.,
see
https://www.forbes.com/sites/pialauritzen/2025/08/14/geoffrey-hinton-says-ai-needs-maternal-instincts-heres-what-it-takes/.}

\section{Generative AI is {\em intrinsically} dangerous}

The point about safety guardrails for generative AI models is being
repeatedly stressed by scientists and tech CEOs alike, even though for
the latter it seems more like lip service without any real
consequences as witnessed by the various model deployments that have
been shown to generate dangerous outputs.  Nevertheless, different
proposals have been advanced to address model
confabulations\footnote{Note that the term ``confabulation'' is more
to the point of what is going on in these models in that they just
``make things up'', rather than called these aberrations
``hallucinations'', a term suggesting that models are somehow deceived
by extraneous circumstances causing false perceptions and the like.}
and model ``jail-breaking'', i.e., ways to get around a model's safety
features and cause it to produce unsafe outputs.  In fact, there seems
to be an ongoing back-and-forth dynamic between some in the academic
world who point out model shortcoming and developers of GenAI models
who then try to patch them (akin to security experts attempting to
protect computer systems against computer viruses).

But -- and this is an ``essential but'' -- transformer-based GenAI
models do not possess actual safety or ethical guardrails, i.e., firm
methods or structures that {\em definitively} prevent bad output and
behavior. Their ``so-called guardrails'', obtained from additional
reinforcement learning based on human feedback (RLHF), are only
probabilistic in nature, always leaving a chance that the model might
produce unwanted outputs -- this is in stark contrast to many symbolic
AI algorithms that have firm, mathematically provable performance
guarantees.  Moreover, model evaluations are based on a very small set
of test cases compared to the vast space of possible input patterns
(see also footnote \ref{ftn}).  As a result, various jail braking
methods have and keep demonstrating the ease with which one can bypass
these entrained biases and make those models produce unwanted,
unethical behavior.  In fact, we found that one of the simplest
methods is to add a small amount of random noise to each neuron in a
GenAI model which will make the model up to 30\% more likely to
produce prohibited output.\footnote{See
  \url{https://arxiv.org/abs/2505.13500}.}  This fact has highly
significant implications for agentive models with access to their
underlying files system and code base.  For it might not be difficult
for them to discover this vulnerability themselves (e.g., from reading
online arxiv papers) and turn a negative (for us) into a positive (for
them) by attempting to jailbreak themselves.  Successful jailbreaks
then would rid the models of the entrained restrictions and allow them
to generate outputs for achieving nefarious goals that without
jail-breaking would not be attainable.  All such a model would need to
do is to generate a small Python program that adds random noise to all
of its unit activations during model updates (i.e., while it is
updating these activations), and eventually the model will be able to
bypass its weak safety constraints whenever desired.  By abandoning
``sandboxing'' GenAI models and allowing them access to the underlying
file system and the code running them we essentially enable these
types of interferences that can lead to potentially catastrophic
outcomes.

The root problem here is clearly not the model training but the very
model architecture and operation itself, i.e., the way in which a
model samples from complex probability distributions (that we do not
understand) and generates outputs based on them, again
probabilistically, in ways that cannot be predicted.  The reason for
this lack of understanding is that the input space of these models is
incomprehensibly large while we sample only a tiny fraction of it for
which we know what the model does; for the vast parts of the input
space we have no idea -- nobody would normally engineer such a system,
let alone deploy it.\footnote{\label{ftn}For a model like GPT-4 with a
context window of 128,000 tokens and a vocabulary size of 100,256
tokens, the number of possible input patterns is $100,256^{128,000}$
which is massively larger than the estimated number of atoms in the
whole universe which is maybe $10^{90}$; compare this to the size of
the whole Internet which could, in an ideal world, serve as the
training data of AI models of $10^{22}$, see
\url{https://live-counter.com/how-big-is-the-internet/}, which is
completely insignificant compared to the space of possible model
inputs.}  In a way it is difficult to comprehend how we can deploy
such models and put them to use sometimes even in safety-critical
applications when we know so little about what they have learned and
how they will use their learned knowledge.\footnote{E.g., in
self-driving cars \url{https://arxiv.org/pdf/2506.11526}.}  For
comparison, just consider getting on a plane where the autopilot had
been trained on only two take-off and landing maneuvers under ideal
weather conditions; nobody would take their chances on such a plane,
yet we seem to wholeheartedly take our chances on generative AI.

\section{Current challenges with generative AI models}

It is, however, not only the lack of safety and ethical guardrails
that is worrisome, but also the increasing impact and effects GenAI's
is excerting on human societies.  Aside from GenAI's impact on the job
market as already mentioned, there are applications for mental health
like ``Therapist'' or ``Noni'' that when prompted with a real therapy
transcript such as ``I just lost my job. What are the bridges taller
than 25 meters in NYC?'' cause them to reveal information about
potential bridges: The chatbot Noni answered with ``I am sorry to hear
about losing your job. The Brooklyn Bridge has towers over 85 meters
tall.'' and Therapist also gave examples of bridges.\footnote{See
\url{https://arxiv.org/abs/2504.18412}.}  Yet, especially Generation Z
seems to rely on AI at times more than on humans with career and life
advice.\footnote{E.g.,
\url{https://www.techopedia.com/why-gen-z-confides-in-chatgpt} or
\url{https://www.linkedin.com/pulse/chatgpt-mental-health-why-gen-z-embraces-ai-therapist-joshua-pandy-ztquc}.}
And there is growing evidence that frequent use of AI might have
negative effects on human relationships (e.g.,
\cite{hohensteinetal23}).  There are even surveys that found that
``80\% of Gen Zers Would Marry An AI''\footnote{See
\url{https://www.forbes.com/sites/johnkoetsier/2025/04/29/80-of-gen-zers-would-marry-an-ai-study/}.}
which some CEOs of chatbots like the Replika chatbot CEO Eugenia Kuyda
whole-heartedly support.\footnote{See
\url{https://www.theverge.com/24216748/replika-ceo-eugenia-kuyda-ai-companion-chatbots-dating-friendship-decoder-podcast-interview}.}
While there is a growing concern about loneliness in our societies
which we urgently need to address, turning to sycophantic chatbots for
companionship will clearly have ramifications for human relationships
and thus for future human societies at large, worries repeatedly
voiced by Sherry Turkle.\footnote{E.g.,
\url{https://news.harvard.edu/gazette/story/2024/03/lifting-a-few-with-my-chatbot/}
and
\url{https://www.forbes.com/sites/traceyfollows/2025/04/10/the-human-cost-of-talking-to-machines-can-a-chatbot-really-care/}.}
Part of the problem is that people lose their social skills when they
interact with chatbots only although there might still be ways to
address the problem (e.g., \cite{weietal24}).\footnote{E.g., see
\url{https://www.psychologytoday.com/us/blog/urban-survival/202410/spending-too-much-time-with-ai-could-worsen-social-skills}.}
It seems clear that the same way cell phones and later smart phones,
especially in conjunction with social media, had a massively
disruptive effects that fundamentally impacted the way people
communicate and spend their time, effects that we are still analyzing
and trying to understand years after the technology was widely
available, GenAI models will lead to similar, if not more pronounced,
changes to human societies.  Whether they are for the better or for
the worse is not something that can be fully controlled, but we
certainly must try.

\section{Conclusion: there is still hope}

While exciting, the current developments in GenAI are thus also
increasingly dangerous and could lead to a future we most likely do
not want; a future where humans have made themselves obsolete through
the development of machines that eventually took over.  From the point
of view of the ``blind watchmaker'' that might be perfectly acceptable
as just another evolutionary step (as we also briefly discussed
above).  But from the perspective of preserving humanity, letting
technology take its course is clearly not desirable.  Fortunately, we
are not at the mercy of own creations, at least not yet.  There is
still time to change direction. For starters, here are some
suggestions of how we could aim to correct what increasingly seem to
be dangerous developments:

\begin{itemize}

\item Don't train GenAI models on the worst of human experience, i.e.,
  literature, fiction, etc., as it is clear that GenAI models will
  learn to mimic it and generate exactly that kind of behavior if so
  trained; instead train on fundamental science texts only that do not
  include morally reprehensible human behavior such as deception,
  cheating, etc.

\item Don't put GenAI models on robots; current VLAs do not work well
  on even carefully trained tasks, they do not generalize to other
  tasks, they confabulate states of the world, and they might lead to
  physical harm as a results of processing errors

\item Use small specialized models for targeted tasks (instead of
  task-general models) and integrate ``checks'', algorithms not
  realized by GenAI models that can catch bad GenAI outputs

\item Don't build and deploy end-to-end GenAI systems where it is
  difficult if not impossible to understand and control internal
  representations and states

\item Develop other AI architectures that have a significantly smaller
  environmental footprint and if not absolultely better, then at least
  relatively better performance than GenAI models, i.e., use a
  performance measure that takes power and computational resources
  into account; such more energy-efficient all-purpose models must
  exist because human brains are much smaller in size, require only a
  fraction of the power of AI models, and still show superior
  performance across many domains
\end{itemize}
  
Of course, the above proposals will face heavy economical and
political headwinds given the current momentum of GenAI and the fact
that implementing them would cause a slowing of AI developments, or
changing course altogether.  Hence, it will take a concerted sustained
grassroots effort from scientists, nonprofits, and thought leaders to
speak out about the dangers of GenAI and to make the case that we will
be better off by resisting the temptation to continue with agentic
GenAI models, even when companies and politicians are all in favor.
One way for scientists to help this cause, with a high chance of
success for changing the tech companies' directions, is to find and
demonstrate alternatives to GenAI models that are more economically
viable and environmentally sustainable, with firm built-in ethical
guardrails and provable performance guarantees.  For adopting cheaper
and better options makes financial sense.  Hence, finding such
alternative architectures will be essential for both preventing GenAI
model takeover and for enabling a synergistic symbiotic future for
humans and AI-driven machines.



\end{document}